# Design and Implementation of Multistage Interconnection Networks for SoC Networks

Mahsa Moazez[1], Farshad Safaei[2], Majid Rezazadeh[2]

[1]Department of Electrical, Computer and IT Engineering, QIAU, Qazvin, IRAN
[2] Faculty of ECE, Shahid Beheshti University G.C., Evin 1983963113, Tehran, IRAN
m.moazez@qiau.ac.ir, f_safaei@sbu.ac.ir, m.rezazadeh@mail.sbu.ac.ir

## Abstract

*In this paper the focus is on a family of Interconnection Networks (INs) known as Multistage Interconnection Networks (MINs). When it is exploited in Network-on-Chip (NoC) architecture designs, smaller circuit area, lower power consumption, less junctions and broader bandwidth can be achieved. Each MIN can be considered as an alternative for an NoC architecture design for its simple topology and easy scalability with low degree. This paper includes two major contributions. First, it compares the performance of seven prominent MINs (i.e. Omega, Butterfly, Flattened Butterfly, Flattened Baseline, Generalized Cube, Beneš and Clos networks) based on 45nm-CMOS technology and under different types of Synthetic and Trace-driven workloads. Second, a network called Meta-Flattened Network (MFN), was introduced that can decrease the blocking probability by means of reduction the number of hops and increase the intermediate paths between stages. This is also led into significant decrease in power consumption.*

## 1. INTRODUCTION

Multiprocessor systems are the only way to achieve high signal processing. The performance evaluation of such systems is dependent on the number of system processors and the access time of each processor to the processing unit. The processors get access to the memory unit through an interconnection network. Multistage interconnection networks (MINs) are a novel approach to implement connections among processors and memory modules. In fact, MINs assign available resources to network components efficiently and cause appropriate trade-off between performance and cost in Networks-on-Chip (NoCs). Furthermore, the bandwidth division is made in the best possible form among different partitions of a MIN with regard to links connectivity [1]. Indeed, a good interconnection design for processors is a key point to evaluate the performance of a system. For example, injection of uniform workload is responded by a linear increment in assigned bandwidth and logarithmic increment in latency proportional to number of nodes [7]. Therefore, a multiprocessor system could be analyzed and evaluated as an NoC using

MINs [3, 4]. Considering availability of paths to establish new connections, MINs are classified into three categories: blocking, non-blocking, and re-arrangeable networks [2, 8, 9].

The most obvious problem of MINs is the blocking problem and impossibility of the implementation of appropriate routing algorithms since there is only a unique path between every input-output pair. A connection between a free input-output pair is not always available because of probable conflicts between the existing connections. Hence, in this paper, a novel structure which referred to as Meta-Flattened Network (MF) is introduced in order to increase the number of paths between every pair of sources and destinations. By using this structure we can reduce the likelihood of the blockage using different routing algorithms.

The paper is structured as four major sections. In the first section, MINs are briefly introduced. Then, the main idea of Dally's flattened network is proposed [10]. MF-MINs are presented in Section 3. Finally, the performance of MF-MINs and the conventional MINs are compared in terms of three parameters; i.e., the power consumption, the message latency, and the network throughput under both Trace-driven and Synthetic workloads.

## 2. DELTA NETWORKS

Delta networks were proposed by Patel [10, 11] as an inexpensive alternative for crossbars. They are composed of sub-networks called Banyan. It is a kind of blocking networks which have self-routing property. Therefore, a Delta network can be viewed as a fundamental topology for *Omega*, *Baseline*, *Butterfly*, and *Generalized-cube* networks structure [1, 2]. Figure 1 illustrates two popular structures of delta networks.

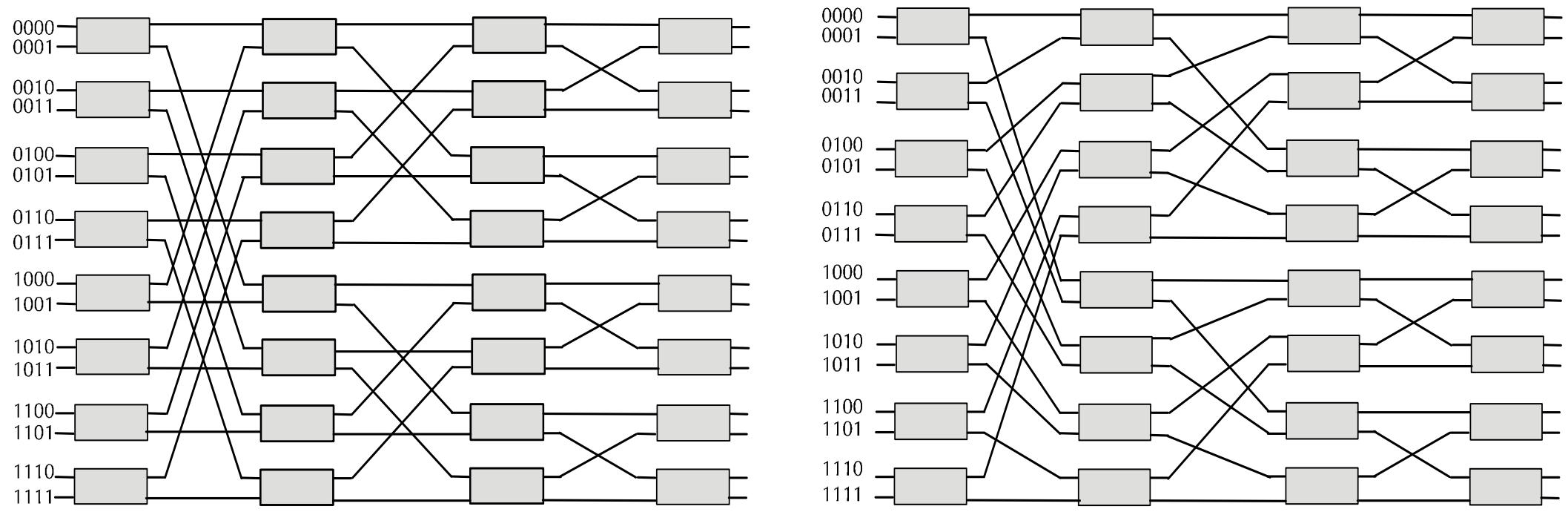

(a) (b)

Figure 1. (a) Butterfly network; (b) Baseline network

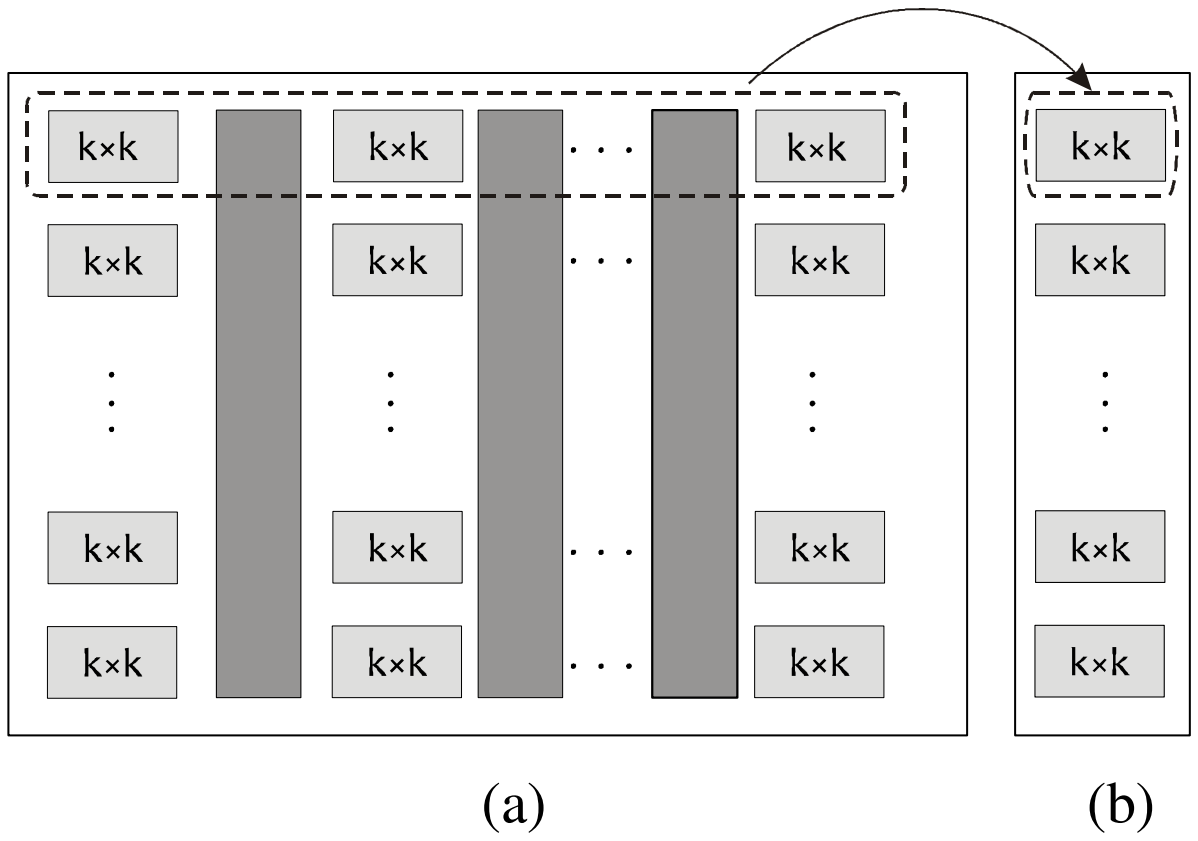

Figure 2. (a) A generic schema of Delta networks; (b) Flattened Delta network

## 3. META FLATTENED DELTA NETWORK AS AN APPROACH FOR IMPROVEMENT OF PERFORMANCE

There is no path diversity in Delta networks, which leads to poor performance for inconsistence workloads. Reduction in the number of stages decreases the number of possible places for conflict, which consequently brings about increment of the throughput. So, the flattened structure is proposed for MINs [12, 13]. To create a flattened structure, all of the routers in each level of a MIN are merged into one router (see Figure 2). This structure reduces the number of hops between the source and destination nodes. Further, it might reduce the probability of blocking to zero with the implementation of different types of the routing algorithms.

### 3.1. Meta Flattened Delta Network (MF)

As it was discussed in the previous sections, MINs facilitated the passage into parallel processing, but a big challenge in these networks relates to no implementation of routing algorithms as a result of a unique path in a pair of source and destination nodes. Moreover, a flattened structure better fit high-radix interconnection networks; however, radix growth increases the complexity of implementation of a flattened structure exponentially. Since the number of routers, inputs and outputs and control signals increase, more area is occupied and the complexity of routing algorithms increases. Hence, the idea of Meta-Flattened (MF) structure for on-chip interconnection networks presents in order to create parallel processing. Using this structure, the blocking problem is solved fairly in MINs and the number of hops among sources and destinations is decreased too. In addition, MF structure is less complex than flattened one and occupies smaller area.

#### 3.1.1. The sketch

In MF networks, the structure of the first and the last stages remains constant. Also, similar to a flattened network, the intermediate stages are merged to form a single stage. Two methods can be adopted to flatten intermediate stages. First, the stages can be flattened in groups of two, which is mainly applicable to networks with the even number of stages. Second, all intermediate stages can be implemented as a flattened network. In the first method, the router degree is low and it can be implemented more easily than the second one. In this case, the network structure is more similar to a MIN rather than a flattened network. In the second approach, the structure is analogous to the structure of a flattened network. The degree of intermediate stages increases and the blocking problem mitigates partly comparing with the first method.

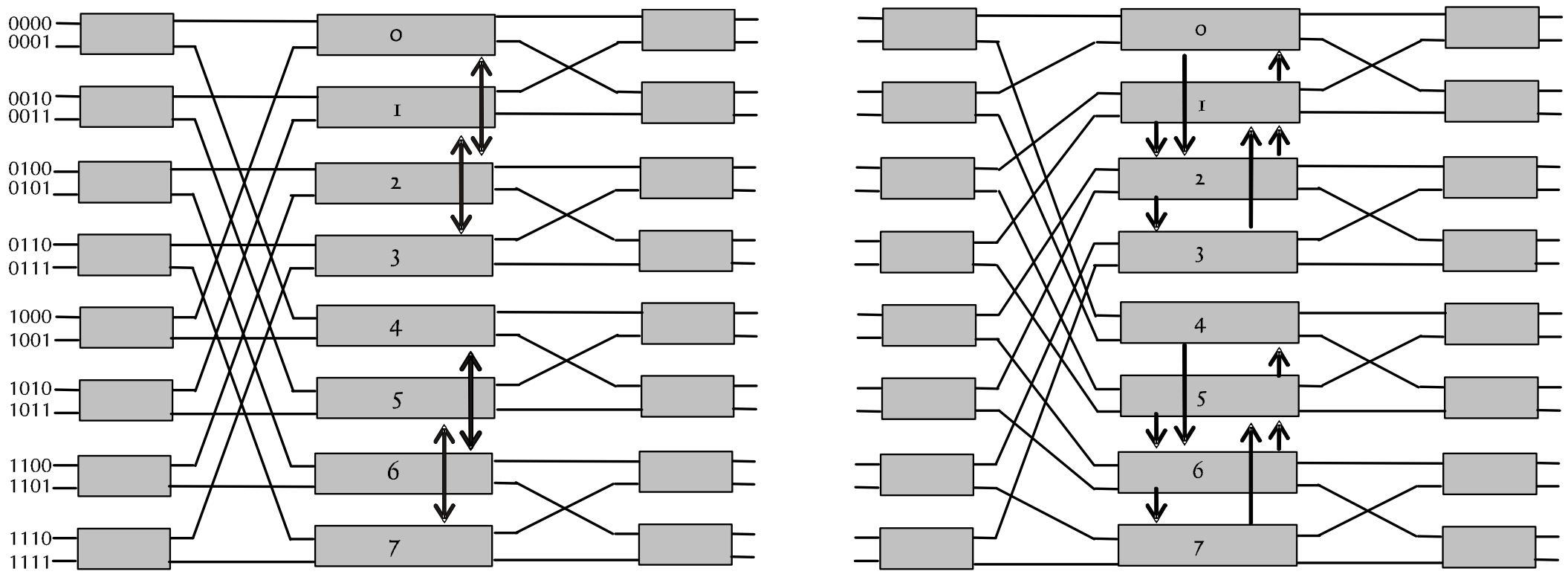

(a) (b)

Figure 3. (a) MF-Butterfly; (b) MF-Baseline

### 3.1.2. MF-Butterfly and MF-Baseline networks

Figures 3(a) and 3(b) represent the organizations of MF-Butterfly and MF-Baseline with 16 inputs, respectively. In these structures, the second and third stages of Butterfly and Baseline networks in Figure 1 are merged together and are flattened while the first and the last stages remain unchanged. As shown in this figure, the number of stages is reduced to three. Moreover, the number of inputs and outputs in the intermediate routers increases.

## 4. Simulation and Results

Verilog language is used to simulate the proposed structures in this section. Synopsys DC is utilized to synthesis on-chip with 45nm Nangate technology. All of the designs are simulated at 1.1V supply voltage using Modelsim 6.5b with 16nm CMOS technology. In this simulation, the following assumptions are made [1, 3-5]:

- There are 32 processors as inputs of the networks
- *Wormhole switching* [1] is used
- The number of the physical channels and the number of inputs are the same
- The networks performance is compared to each other under six workloads including three Synthetic [14] (i.e. uniform, exponential, and normal) and three splash Trace-driven workloads [15] (i.e. FFT, Water-Squared, and Water-Spatial) for all MINs (i.e. Omega, Butterfly, Baseline, Generalized Cube, Beneš, and Clos). Moreover, we reported the performance merits of suggested MF networks (i.e. MF-Baseline and MF-Butterfly) compare with the traditional MINs
- The size of each message is supposed to 2 flits under Synthetic workloads
- In MF Networks, the applied routing algorithm is fortified with the *adaptive routing* [1]

## 4.1. The power consumption

According to Table 1, the power consumption of networks is different. Omega, Butterfly, Baseline and Generalized-cube networks have almost the same power consumption. The difference in their values does not exceed a ten-thousandth of milli Watts (mW) because these networks have equal number of routers and wires. The only difference among these networks relates to the permutation of connections among routers. For the same reason, Beneš and Clos networks have approximately equal power consumption. This is true for MF-Baseline and MF-Butterfly networks too. Further, the table illustrates that MF networks have less power consumption compared with conventional Baseline or Butterfly in order to decrease in the stages despite of the fact that intermediate routers are getting larger.

Table 1. The power consumption of the networks

| Network | Power Consumption (mw) |
|---|---|
| Omega | 168.55 |
| Baseline | 168.58 |
| Butterfly | 168.6 |
| General-cube | 168.6 |
| Beneš | 300.11 |
| Clos | 300.09 |
| Meta-Flattened Baseline | 149.65 |
| Meta-Flattened Butterfly | 149.66 |

According to the results, the power consumption of MF MINs has 13% improvement versus MINs ones.

## 4.2. Message Latency

The network workload refers to the pattern of traffic which is applied at the network terminals over the time. Understanding and modeling the load is led to design and evaluate networks and routing functions [1]. Figure 4 shows the abundance of FFT workload. In this figure, horizontal and vertical axes indicates the node number and the amount of usage that node in inputs and outputs, respectively. As seen in the figure, nodes 15 and 24 tolerate the most traffic in inputs. On the other hand, nodes 0 to 3 are most used as the destination of messages. Hence, buffers of these routers become more congested and this leads to increase in the total message latency.

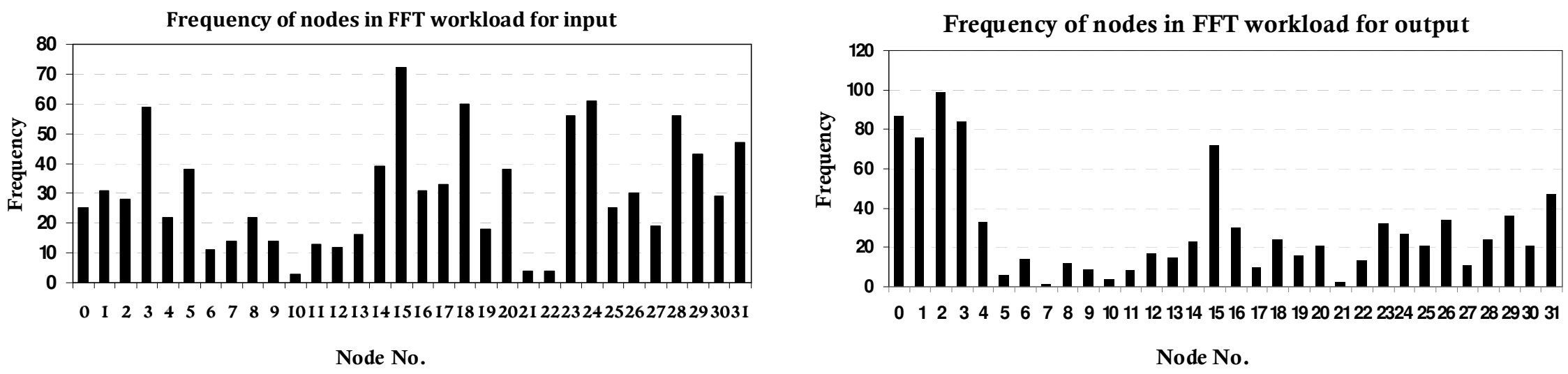

(a) (b)

Figure 4. The frequency of the nodes in FFT workload; (a) input nodes; (b) output nodes

Figure 5 shows the traffic distribution over the inputs and outputs for Water-Nsquared workload. Due to the traffic distribution, nodes 16 and 23 are most selected nodes as the source and destination of messages while the others have the same conditions.

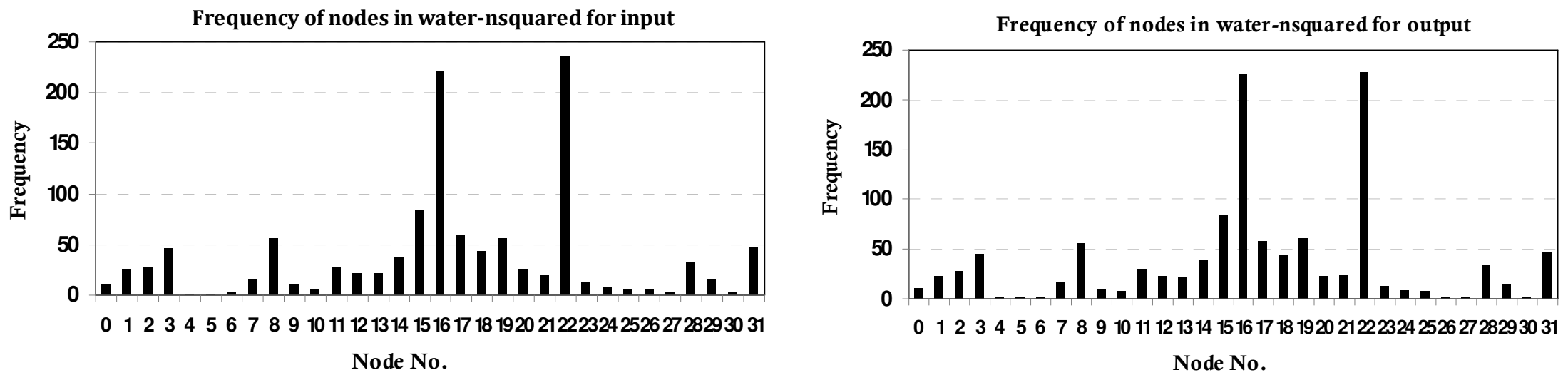

(a) (b)

Figure 5. The frequency of the nodes in Water-Nsquared workload; (a) input nodes; (b) output nodes

Figure 6 represents the traffic distribution over the inputs and outputs for exponential workload. As we can see, this traffic is distributed equally among all the nodes throughout the networks. The message latency is increased due to increase the number of messages with high hop count.

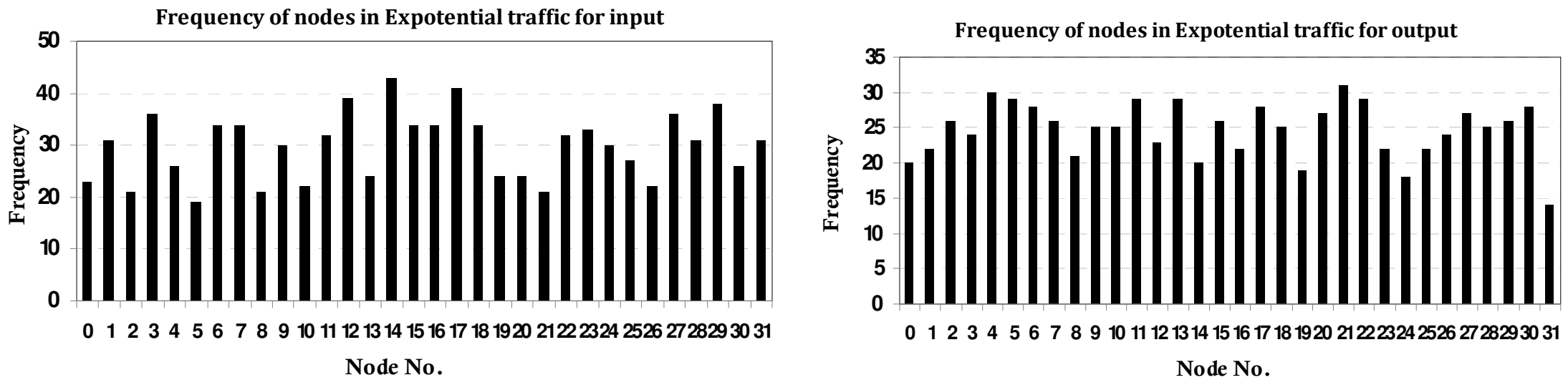

(a) (b)

Figure 6. The frequency of the nodes in Exponential workload; (a) input nodes; (b) output nodes

Figures 7 to 10 exhibit the network latency under different workloads. The latency for various networks operating under different workloads is of different values. For example, in MF-Baseline network, there are routers with two and six I/O numbers and this difference in the number of I/Os is an important factor which determines the latency of the network under various types of workloads. MF-Butterfly network is similar to MF-Baseline network, but all intermediate routers are similar to each other and have the same number of inputs and outputs.

In Figure 7, the message latency for different networks under Trace-driven workloads are demonstrated. It shows the average message latency for FFT, Water-Nsquared, and Water-Spatial workloads. Under FFT workload, the latency of MF networks is near to the conventional MINs. As we addressed (see Figure 4), the frequency of nodes 2, 15, 23, 24, 26 is more than the other nodes under this workload. The nodes belong to the routers with minimum inputs and outputs which tolerate the maximum traffic; hence the flits of the message must expend more time to cross over them and it leads to increase of total message latency. So, compared to the traditional MINs a significant improvement in the latency is perceived for MF networks.

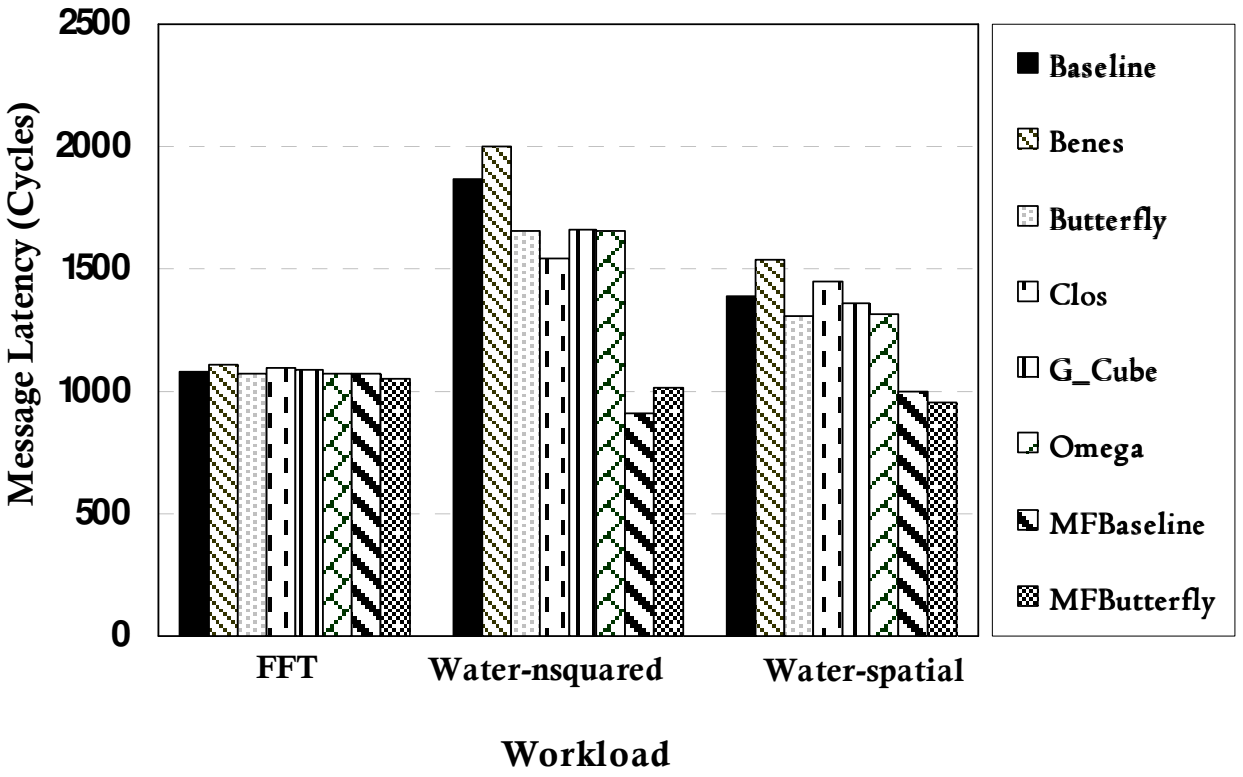

Figure 7. The message latency of the networks under Trace-driven workloads

Figure 8 demonstrates the message latency under Exponential workload. MF-Butterfly network has the minimum latency and Beneš network has the maximum latency in order to have several stages among the networks. The reason of the proper operation of MF networks under this workload is related to usage of the same nodes in inputs and outputs. This phenomenon reflects the impact of the several paths for inputs and outputs and improvement of MF networks operation.

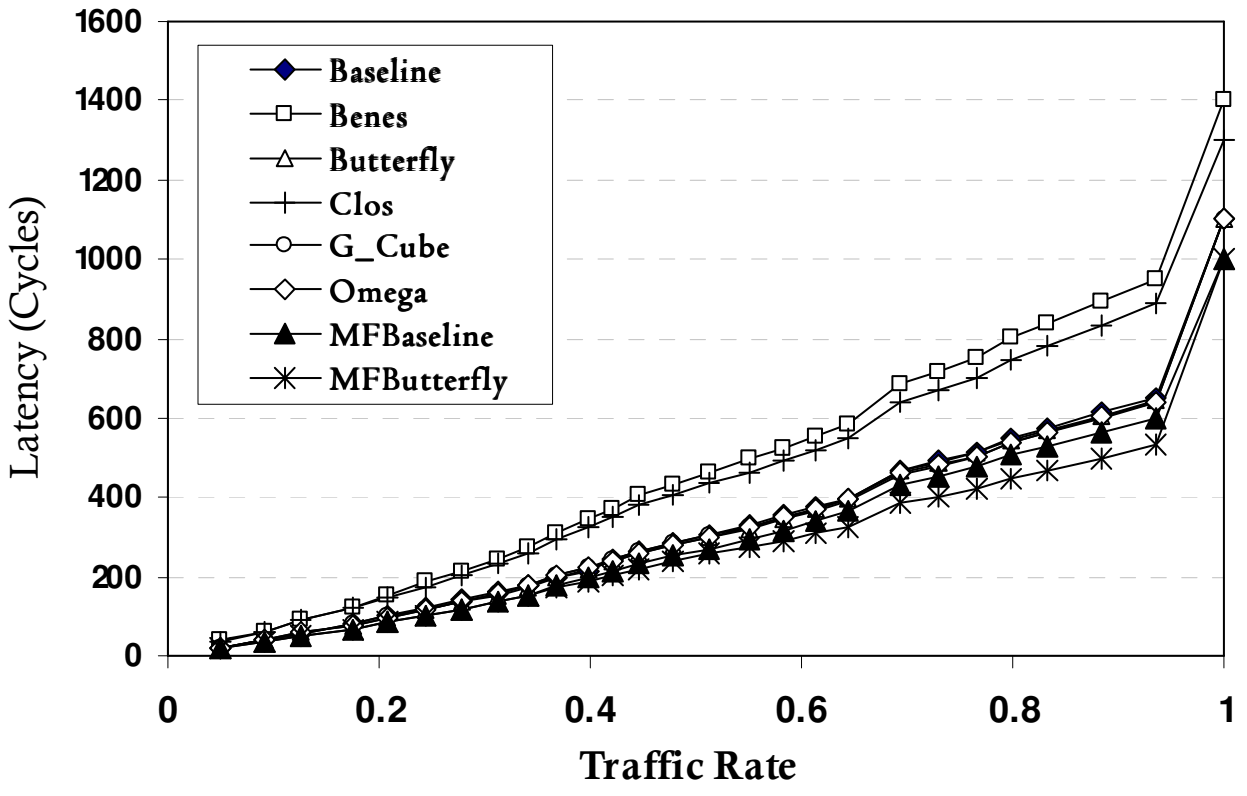

Figure 8. The message latency of the networks under Exponential workload

Similarly, Figure 9 exhibits the message latency under Uniform workload. In this traffic, MF-Baseline has shown minimum latency compared with the other networks. Gradually, when the traffic volume increases the message latency lengthens too, even longer than the conventional MINs. The reason of the better performance refers to the irregular structure of the network and difference in the number of inputs and outputs routers. For example, there are routers with 6 inputs and 4 outputs, also routers with 3 inputs and 3 outputs. Hence, messages must wait in queues to traverse the network. But, routers of MF-Butterfly network have regular structure and the latency improves considerably.

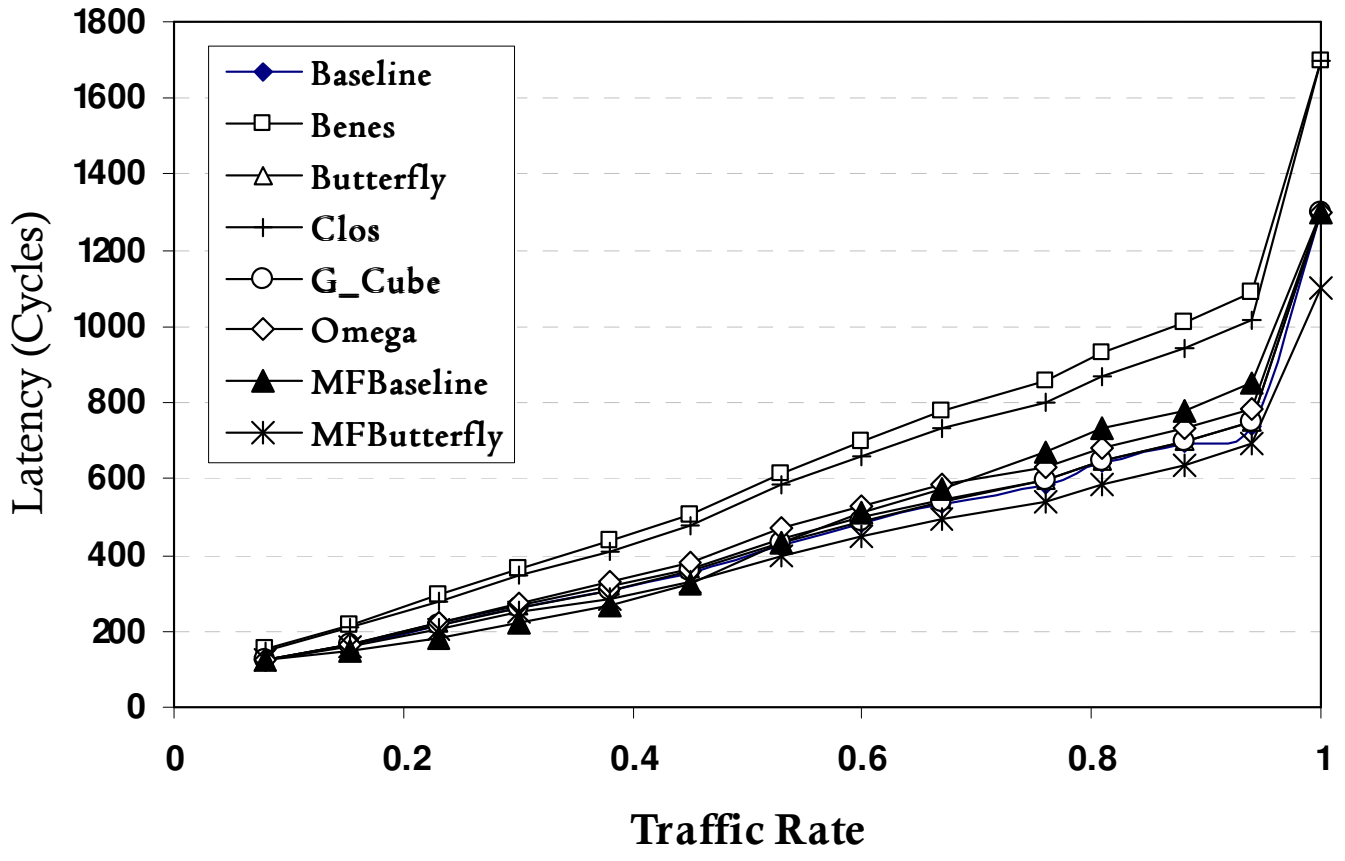

Figure 9. The message latency of the networks under Uniform workload

Figure 10 compares the message latency of MINs under Normal workload. Baseline network has minimum latency among MINs. Despite the same structure of the routers in conventional MINs, network structure has a key role on the performance parameters. Moreover, MF networks have little difference in the performance compared with the conventional MINs because the volume of the traffic is over the nodes which have minimum inputs and outputs. Under this workload, MF-Butterfly network has the best performance because of its regular structure and the path diversity between every input-output pair.

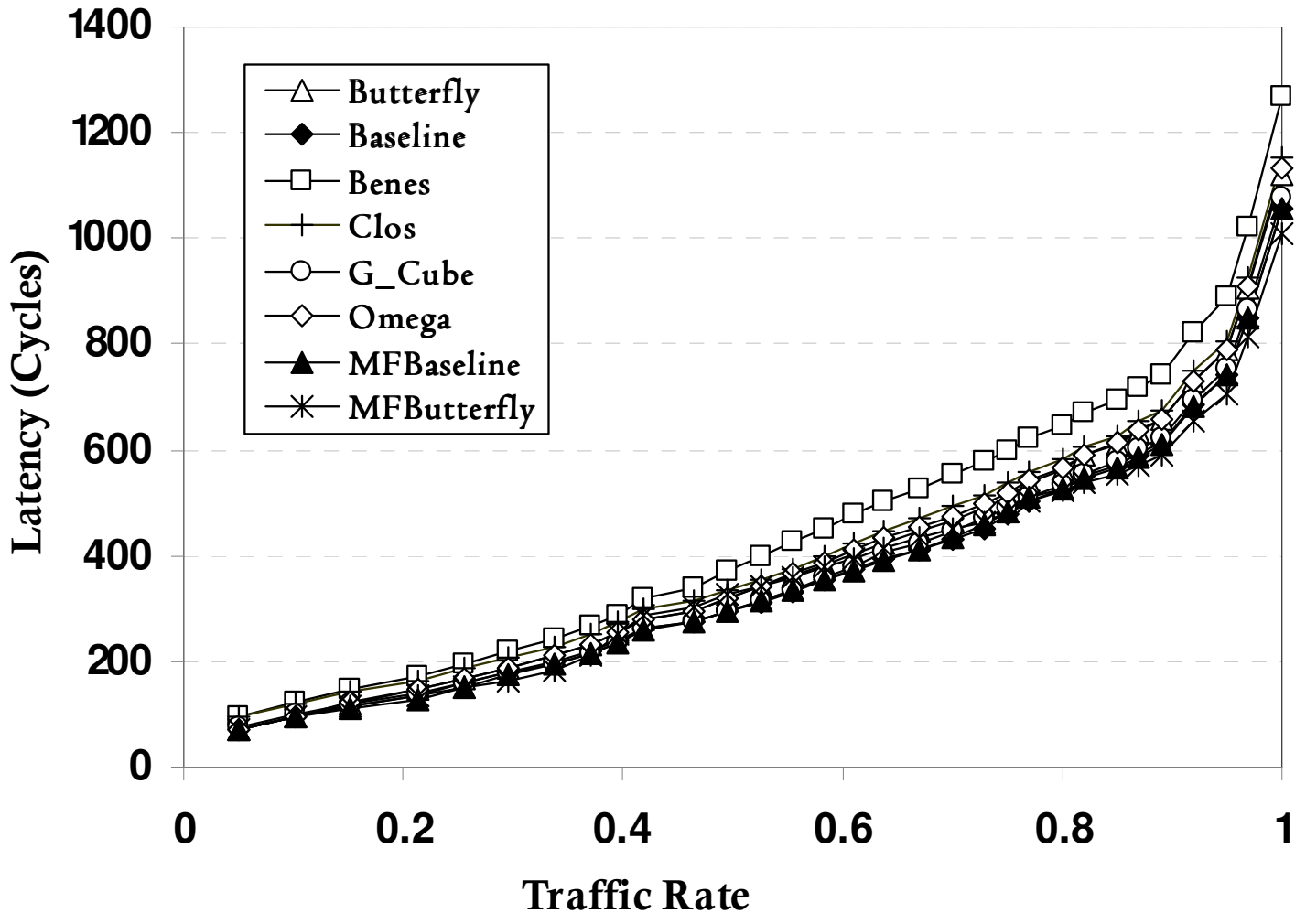

Figure 10. The message latency of the networks under Normal workload

## 4.3. Throughput Evaluation

Figures 11 to 14 show the throughput of different networks under various workloads. In Figure 11, the throughput of Trace-driven workloads is illustrated. In this figure, horizontal axis indicates the type of workload and the vertical axis represents the throughput. As explained in the previous section, under Water-Nsquared workload, MF-Baseline has minimum latency while in Generalized-cube network the same number of flits has been crossed in the longer period of time. Furthermore, under Water-Spatial workload, MF-Butterfly network has increase of 30% in the throughput compared with Butterfly network in order to decrease the number of hops and the number of paths between each input-output pair.

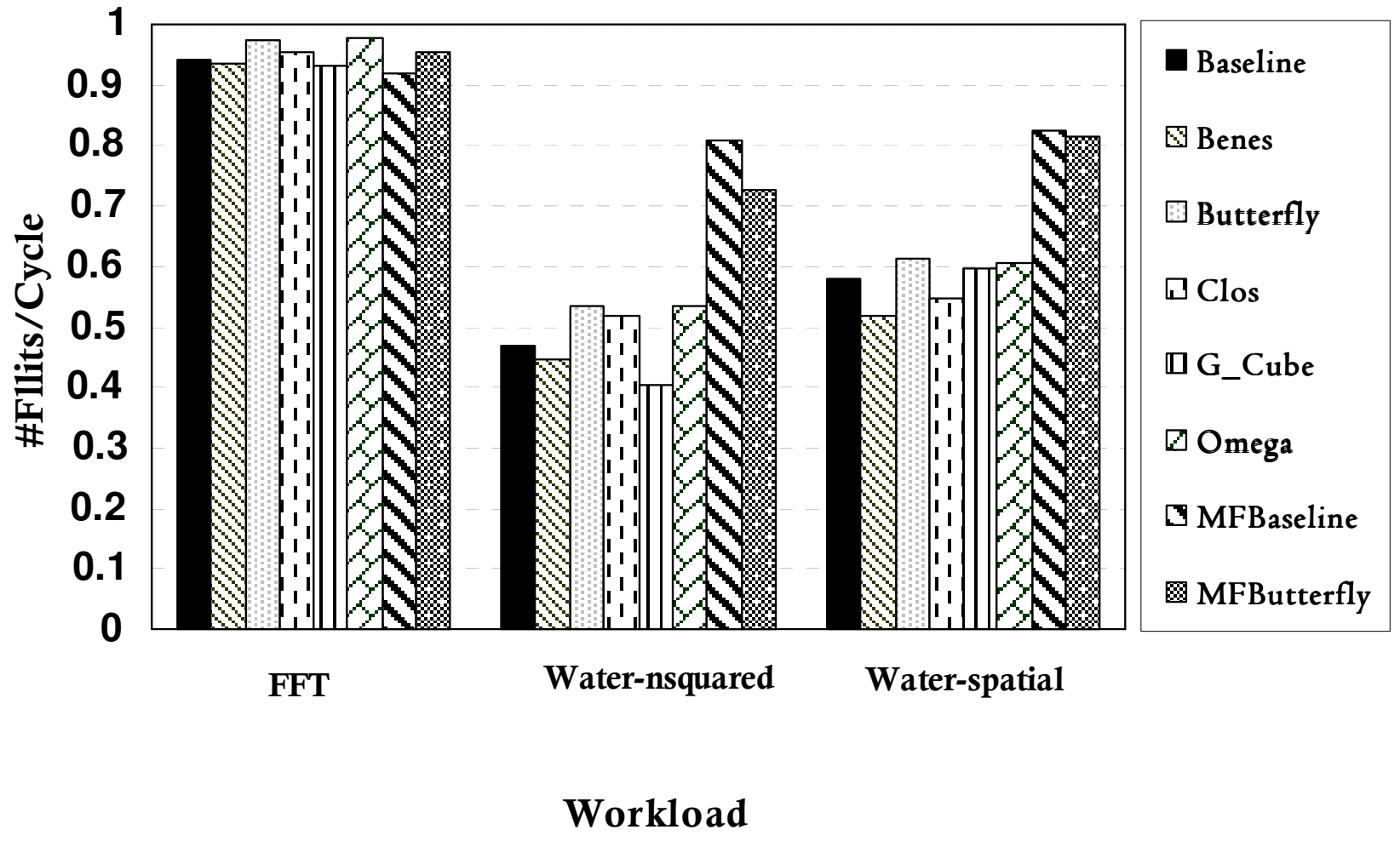

Figure 11. The throughput of the networks under Trace-driven workloads

Figure 12 shows the throughput under Exponential workload. The horizontal axis represents the traffic rate and the vertical axis shows the throughput. The performance of networks under this workload is inversely proportional to the message latency; In other words, the network which has the lowest latency shows a greater throughput. As can be seen in this figure, for traffic rate of 0.4, the throughput increases with a lower slope, then it increases with an appropriate slope. The reason of such behavior refers to increase of the traffic volume that causes the routers can tolerate larger amount of traffic and the messages have to wait more cycles in the queues.

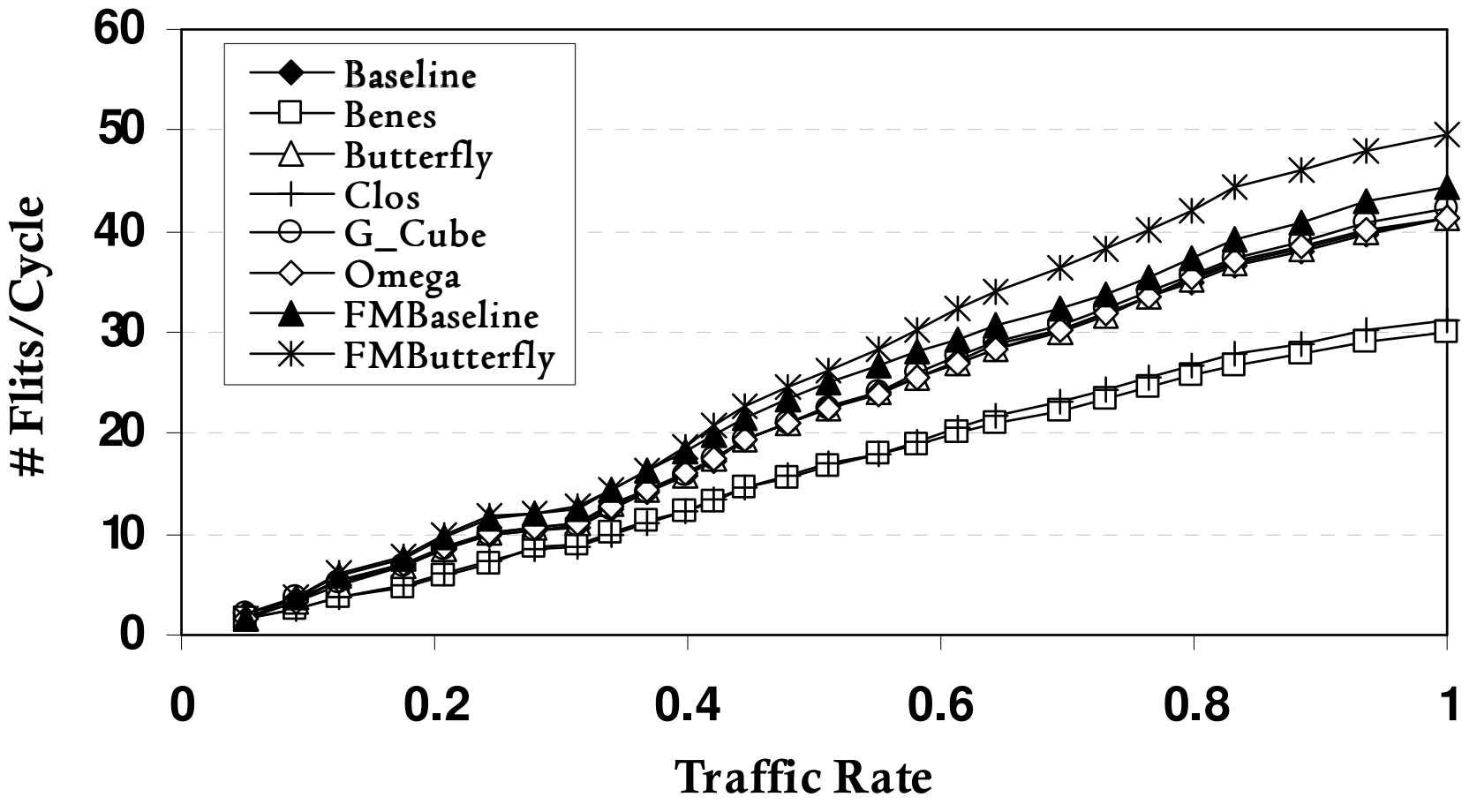

Figure 12. The throughput of the networks under Exponential workload

Figure 13 represents the impact of Uniform workload on the throughput of the networks. In this figure, MF-Baseline has the maximum throughput under the light traffic region, but gradually comes close to Omega network because of the unbalanced structure.

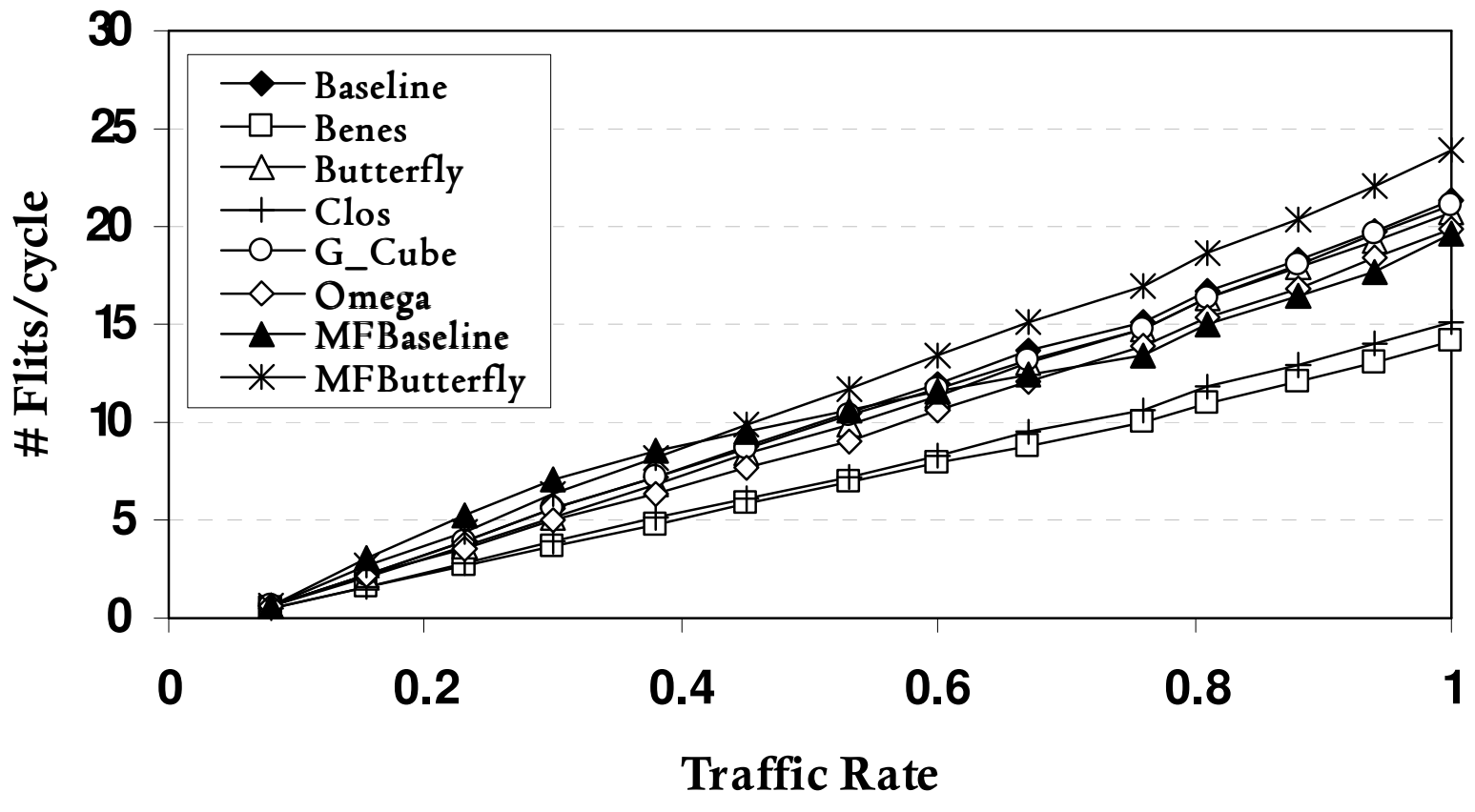

Figure 13. The throughput of the networks under Uniform workload

Figure 14 shows the throughput under Normal workload. Since the number of stages in Clos and Beneš networks is almost double compared to the other networks, the messages must traverse more hops. Thus, the number of messages passing through these networks is less than the other ones. Among the MF networks, MF-Butterfly network shows the better performance and passes more messages in each cycle because of the regular structure and the same degree of routers.

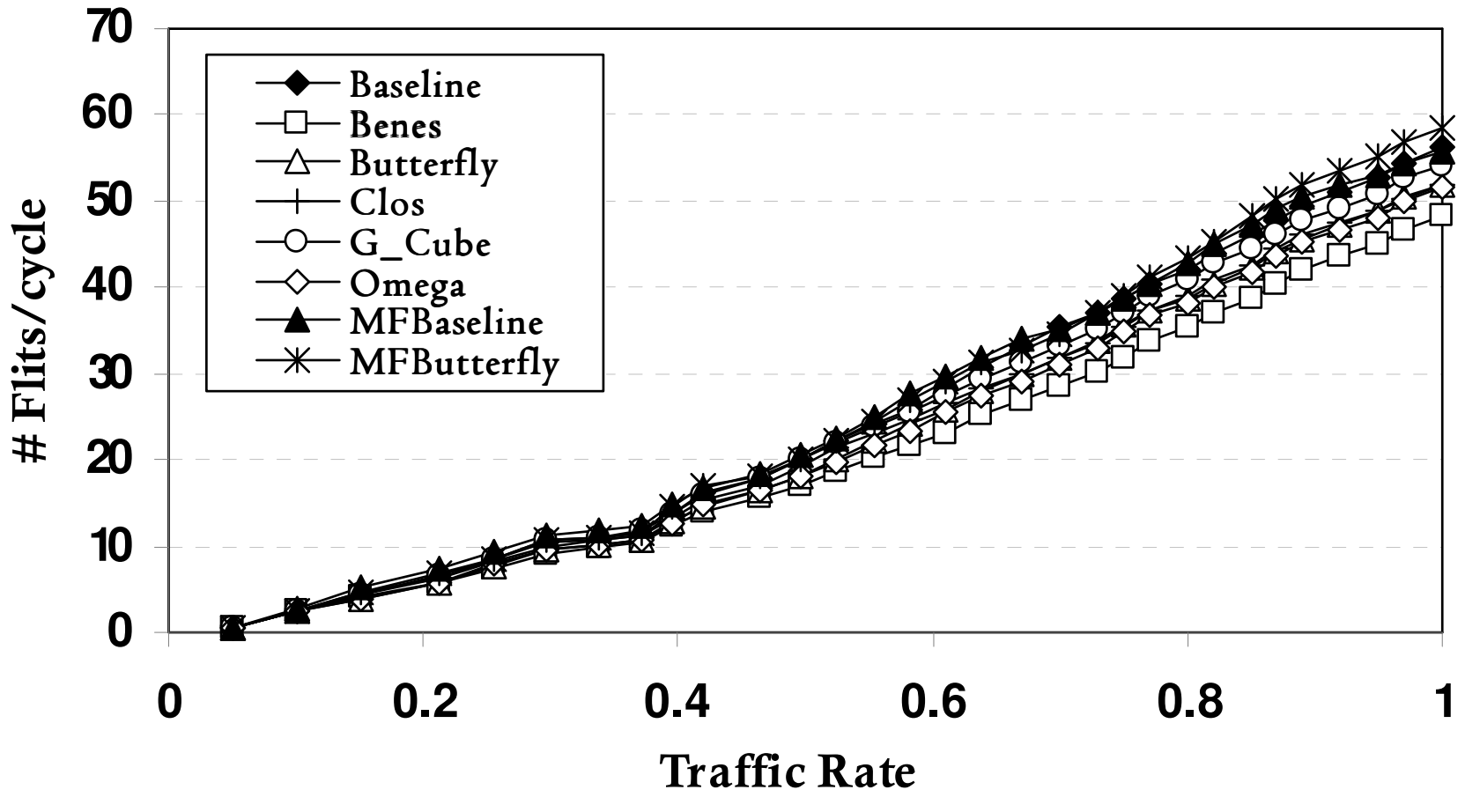

Figure 14. The throughput of the networks under Normal workload

Generally, with reduction of hops in MF MINs, a considerable increase in the throughput and a decrease in the latency are observed compared to the conventional MINs. For instance, an increase of 26% in the throughput is achieved for MF-Butterfly network compared to conventional Butterfly network under Water-Nsquared workload. Furthermore, this value reaches 15% under Exponential workload.

## 5. CONCLUSIONS

One of the fundamental problems in Multi-Stage Interconnection Networks (MINs) is the occurrence of blocking and impossibility of the implementation of appropriate routing algorithms. In this paper, we proposed a novel structure named to Meta-Flattened MIN (MF-MIN) which is able to overcome such problems as well. The suggested structure synthesized with aid of simulation results under Trace-driven and Synthetic workloads. It was shown that the proposed structure is able to largely improve the important performance parameters compared with MINs. The path diversity in each NoC makes it potentially to tolerate faults and failures. However, the occurrence of faults and failures in a network leads to the loss of the global information-carrying ability. This issue provides a much fuller characterization of the vulnerability of the networks. In future work, we will aim to propose metrics to estimate the vulnerability of introduced architecture in the presence of faults.